\documentstyle[aps,floats,graphicx]{revtex}

\let\bi\bf

\begin{document}
\draft \wideabs {
\title{Deconvolution problems in x-ray absorption fine structure}
\author{K V Klementev}
\address{Moscow State Engineering Physics Institute,
115409 Kashirskoe sh. 31, Moscow, Russia}
\date{January 15, 2001}
\maketitle
\begin{abstract}
A Bayesian method application to the deconvolution of EXAFS spectra is
considered. It is shown that for purposes of EXAFS spectroscopy, from the
infinitely large number of Bayesian solutions it is possible to determine an
optimal range of solutions, any one from which is appropriate. Since this
removes the requirement for the uniqueness of solution, it becomes possible
to exclude the instrumental broadening and the lifetime broadening from EXAFS
spectra. In addition, we propose several approaches to the determination of
optimal Bayesian regularization parameter. The Bayesian deconvolution is
compared with the deconvolution which uses the Fourier transform and optimal
Wiener filtering. It is shown that XPS spectra could be in principle used for
extraction of a one-electron absorptance. The amplitude correction factors
obtained after deconvolution are considered and discussed.
\end{abstract}
\pacs{61.10.Ht} }

\section{Introduction}\label{intro}
\begin{figure}[!t]\begin{center}\includegraphics*{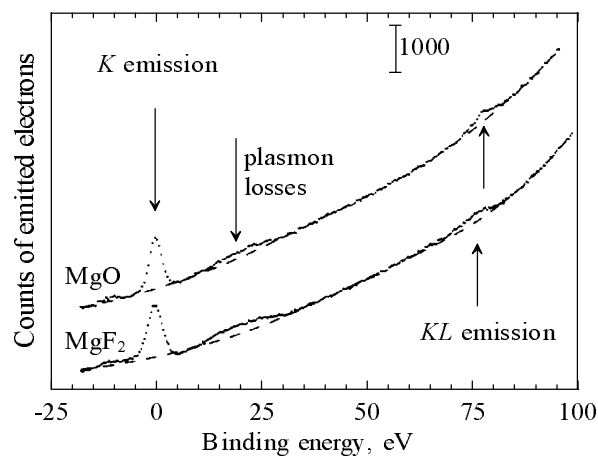}\end{center}
\caption{From \protect\cite{DiCicco94}: XPS spectra of MgO and MgF$_2$ in the
vicinity of the Mg 1$s$ peak. Two secondary structures, due to plasmon losses
and to double-electrton $KL_{2,3}$ excitations, are detected. Zero of energy
is placed at Mg $K$ level ($\sim1300$\,eV).} \label{Cicco}\end{figure}

The chief task of the extended x-ray absorption fine-structure (EXAFS)
spectroscopy, determination of interatomic distances, rms fluctuations in
bond lengths etc., is solved mainly by means of the fitting of parameterized
theoretical curves to experimental ones. However, there exist obstacles for
such a direct comparison: theory limitations and systematic errors. Among
latter are various broadening effects. Fist of all, (i) this is the
broadening arising from the finite energy selectivity of monochromator and
the finite angular size of the x-ray beam. (ii) The absorption even of
strictly monochromatic x-ray irradiation by the electrons of a deep atomic
level gives rise to photoelectrons with the finite energy dispersion due to
the finite natural width of this level and the finite lifetime of the
core-hole. (iii) For x-ray energies far above the absorption edge the process
of photoelectron creation (the outgoing from an absorbing atom) and the
process of its propagation occur for essentially different time intervals. In
other words, just created, the photoelectron `does not know' where and how it
will decay. Therefore the photoionization from the chosen atomic level and
excitation of the remaining system can be considered as independent
processes, and hence the total absorption cross-section, as a probability
density of two independent random processes, is given by the convolution of a
one-electron cross-section and excitation spectrum $W(\Delta E)$. The latter
is the probability density of the energy $\Delta E$ capture at the
electron-hole pair creation and is the quantity measured in x-ray
photoemission spectroscopy (XPS). For light elements there are examples of
such deep and lengthy enough XPS spectra (see figure \ref{Cicco}, taken from
\cite{DiCicco94}).

For all three cases the measured absorption coefficient $\mu_{\rm m}$ is
given by the convolution:
\begin{equation}\label{conv}
\mu_{\rm m}(E)=W*\mu\equiv\int W(E-E')\mu(E')\,{\rm d}E',
\end{equation}
where the broadening profile $W(\Delta E)$ and the meaning of the function
$\mu$ depend on the considered problem. These can be, correspondingly: x-ray
spectral density after the monochromator and the cross-section of ideally
monochromatic irradiation; the Lorentzian function and the cross-section with
a stationary initial level (of zero width); the excitation spectrum and a
one-electron cross-section.

It is the common practice in modern EXAFS spectroscopy to account for the
broadening processes (i)--(iii) at the stage of theoretical calculations by
introducing into the one-electron scattering potential the imaginary
correlation part which represents the average interaction between
photoelectron and hole and their own polarization cloud; in doing so, the
choice of the correlation part is dictated by empiric considerations and can
be different for different systems.

Another approach to the account for the broadening processes is to solve the
integral equation (\ref{conv}) for the unknown $\mu$. To find some solution
of this equation is quite not difficult to do, the simplest way is to use the
theorem about the Fourier transform (FT) of convolution. However, it is known
that the problem of deconvolution is an ill-posed one: it has an unstable
solution or, in other words, the infinitely large number of solutions
specified by different realizations of the noise. Thus, there is evident
necessity for determination of an optimal, in some sense, solution. Yet a
less evident approach exists: to find an appropriate functional of the
solution which itself be stable.

A number of works have been addressed the problem of deconvolution, among
them those concerning the x-ray absorption spectra. Loeffen {\em et al.}
\cite{Loeffen96} applied deconvolution with the Lorentzian function partly
eliminating the core-hole life-time broadening. They used fast FT and the
Wiener filter which is determined from the noise level which, in turn, is
specified by the choice of the limiting FT frequency above which the signal
is supposed to be less than noise. The arbitrariness of such a choice gives
rise to rather different deconvolved spectra, which although remained
obscured in \cite{Loeffen96}.

Recently, for the deconvolution problem with a Lorentzian function Filipponi
\cite{Filipponi00} used the FT and proposed the idea of the decomposition of
an experimental spectrum into the sum of linear contribution, a special
analytic function representing the edge and oscillating part. For the
Lorentzian response, the deconvolution for the first two contribution is
found analytically, for the latter one, numerically. The advance of such a
decomposition is in that fact that now the FT of the oscillating part is not
dominated by the very strong signal of low frequency, therefore the
combination of forward and backward FT gives less numerical errors. Notice
that this method is solely suitable for the analytically given response. In
addition, in \cite{Filipponi00} the choice of the filter function (Gaussian
curve) and its parameterization remained vague. Therefore the issue on the
uniqueness or optimality of the found solution was left open.

In early work \cite{Turchin70eng}, for the solution of ill-posed problems the
statistical approach was proposed. Following that work, in the present paper
we shall consider the deconvolution problem in the framework of Bayesian
method, detailed formalism of which was described in \cite{Klementev01}.
Since the parameterization is naturally involved in the Bayesian method,
there exists a principle possibility to choose an optimal, in some sense,
parameter. Here we shall scrutinize the problem of such a choice which is
relevant to any spectroscopy. We shall show that this problem is absent in
EXAFS spectroscopy because though the EXAFS spectrum itself does depend on
the regularization parameter, its FT does not in the range of real space used
for the analysis. In Sec.~\ref{CompBackgr} we discuss the choice of the
optimal deconvolution for a model Gaussian response and compare the results
of Bayesian approach with the results of FT combined with Wiener filtering.
In Sec.~\ref{instr} we utilize the deconvolution to an experimental spectrum
in order to eliminate the aforementioned broadening processes.

\section{The choice of optimal deconvolution}\label{CompBackgr}

First of all, we shall show that the deconvolution problem really has the
infinitely large number of solutions. In the present paper we use for
examples the absorption spectrum of Nd$_{1.85}$Ce$_{0.15}$CuO$_{4-\delta}$
above Cu $K$ edge collected at 8\,K in transmission mode at LURE (beamline
D-21) using Si(111) monochromator and harmonics rejecting mirror; energy step
$\sim2$\,eV, total amount of points 826 (from 8850\,eV to 10500\,eV), each
one recorded with integration time of 10\,s. Let us take for a while for the
response function a simple model form: $W(E)=C\exp(-E^2/2\Gamma^2)$, where
$C$ normalizes $W$ to unity, $\Gamma$ is chosen to be equal to 4\,eV.

\begin{figure}[!b]\begin{center}\includegraphics*{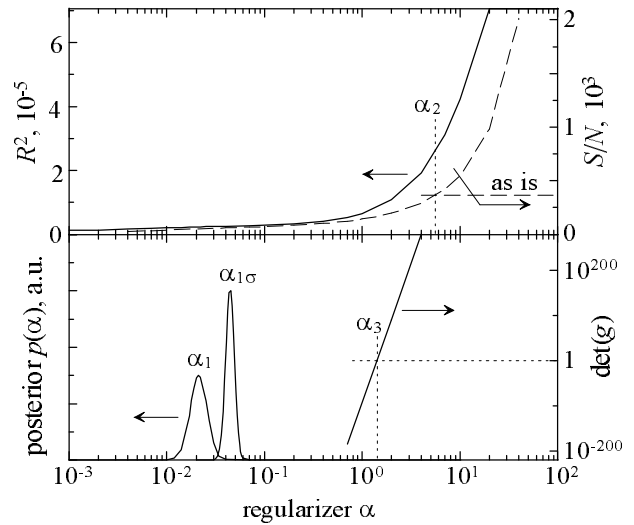}\end{center}
\caption{The quality of the deconvolution $R^2$ vs. regularization parameter
$\alpha$ (solid). Dashed lines --- signal-to-noise ratio before and after
deconvolution. Two peaks --- the posterior density functions
$p(\alpha|\bi{d})$ (left) and $p(\alpha|\bi{d},\sigma)$ (right) for the
regularization parameter $\alpha$. Straight line --- the determinant of
Bayesian matrix as a function of $\alpha$.} \label{regM}\end{figure}

\begin{figure}[!t]\begin{center}\includegraphics*{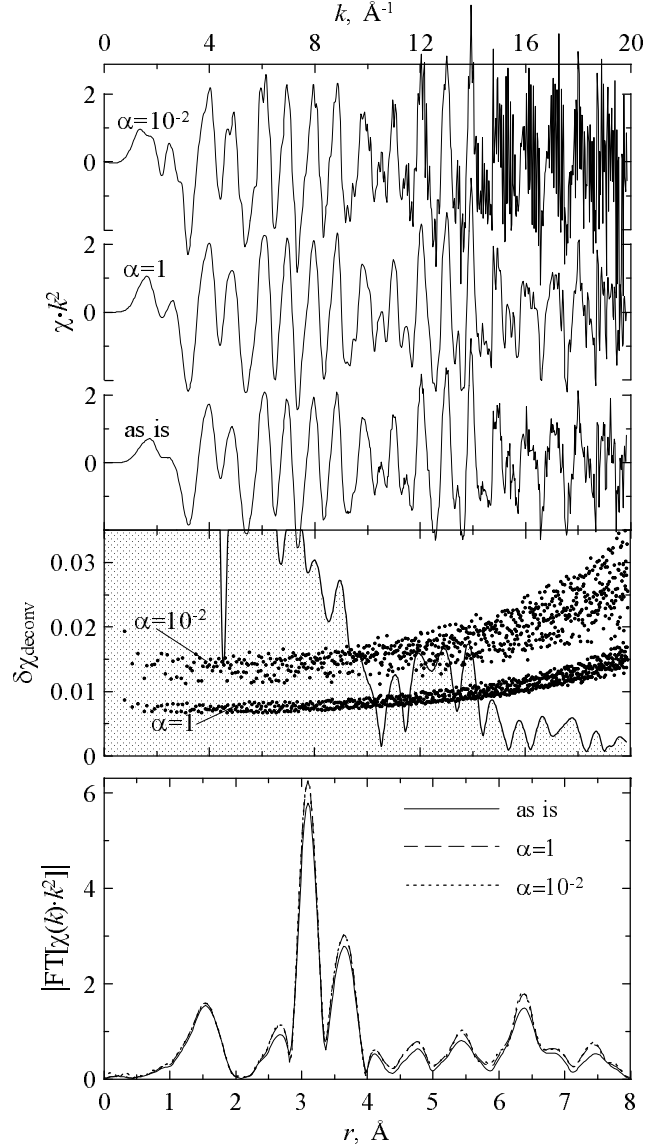}\end{center}
\caption{$\chi\cdot k^2$ obtained without deconvolution and after that with
$\alpha=1$ and $\alpha=0.01$. In the middle --- the envelope of the initial
$\chi$ and rms deviations of the deconvolved values (dots). Below --- the
absolute values of the Fourier transform (the dashed and the dotted curves
practically merge).}\label{regM_FT}\end{figure}

In \cite{Klementev01} we showed {\em how} to construct a regularized solution
of the convolution equation in the framework of the Bayesian approach. For
that, one needs to find eigenvalues and eigenvectors of a special symmetric
$N\times N$ matrix determined by the experimental spectrum, $N$ is the number
of experimental points. Using that approach, find a solution $\mu(E)$ for an
arbitrary regularization parameter $\alpha$ and perform its convolution with
$W$. The $\hat\mu_{\rm m}$ obtained, ideally, must coincide with $\mu_{\rm
m}$. Introduce the characteristics of the solution quality, the normalized
difference of these curves:
\begin{equation}\label{S2}
R^2=\sum_i(\mu_{{\rm m}i}-\hat\mu_{{\rm m}i})^2\Bigr/ \sum_i\mu_{{\rm m}i}^2,
\end{equation}
where the summation is done over all experimental points. Figure \ref{regM}
shows the dependence $R^2$ on $\alpha$. That fact that the quality of the
found solutions is practically the same for all $\alpha\lesssim1$ is a clear
manifestation of ill-posedness of the problem: there is no a unique solution.
How to chose an optimal one? It turns out, that for purposes of EXAFS
spectroscopy there is no need of that and any solution from the optimal range
(here, $\alpha\lesssim1$) is suitable. At {\em arbitrary} $\alpha$ from the
optimal range found the deconvolution $\mu(E)$, extract the EXAFS function
$\chi(k)\cdot k^w$ in a conventional way, where $k$ is the photoelectron wave
number, and find its FT. In figure \ref{regM_FT} we show the EXAFS-functions
obtained after the Bayesian deconvolution with $\alpha=1$ and $\alpha=0.01$,
and their FT's. As seen, although the EXAFS-function itself does depend on
$\alpha$, its FT practically does not. Thus, if one uses for fitting a range
of $r$-space (in our example, up to $r_{\rm max}=8\,$\r{A}) or filtered
$k$-space, the problem of search for the optimal $\alpha$ is not relevant.
Nevertheless, below we propose several approaches to the solution of this
problem, for instance for XANES spectroscopy needs.

(1) For the regularization parameter $\alpha$ itself one can introduce the
posterior probability density function \cite{Turchin70eng,Klementev01} and
choose $\alpha$ with a maximum probability density. It can be done either
using the most probable value of noise or assuming the standard deviation
$\sigma$ of the noise of the absorption coefficient to be known (for our
spectrum $\sigma=9\cdot10^{-4}$, as determined from the FT following
\cite{Newville99}). In figure \ref{regM} these probability densities are
drawn as, correspondingly, $p(\alpha|\bi{d})$ and $p(\alpha|\bi{d},\sigma)$,
and their most probable values are $\alpha_1=0.021$ and
$\alpha_{1\sigma}=0.044$.

(2) The optimal regularization can be determined from the consideration of
the signal-to-noise ratio $S/N$. The Shannon-Hartley theorem states that
$I_{\rm max}=B\ln{(1+S/N)}$, where $I_{\rm max}$ is the maximum information
rate, $B$ is the bandwidth. The authors of \cite{Loeffen96} are of opinion
that deconvolution is a mathematical operation that conserves information.
Therefore from the theorem follows that one pays for an increase in
bandwidth, resulted from deconvolution, via a reduction in $S/N$ ratio. The
thesis on $I_{\rm max}$ conservation is quite questionable, since for
deconvolution one should introduce {\em additional} independent information
about the profile of broadening. What quantity is conserved in deconvolution
is hard to tell. Here to the contrary, for the optimal $\alpha$ we demand to
conserve $S/N$. Define $S/N$ as the ratio of mean values of the EXAFS power
spectrum over two regions, $r<15$\,\r{A} and $15<r<25$\,\r{A}. The
regularization parameter at which the $S/N$ is conserved is denoted in figure
\ref{regM} as $\alpha_2=5.54$. The signal-to-noise ratio can be defined in a
different way. Since the Bayesian methods work in terms of posterior density
functions, for each experimental point one can find not only the mean
deconvolved value but also the standard deviation $\delta\mu_{\rm deconv}$
from which one finds $\delta\chi=\delta\mu_{\rm deconv}/\mu_0$, where $\mu_0$
is the atomic-like absorption coefficient constructed at the stage of EXAFS
function extraction. It is reasonable to compare $\delta\chi$ values with the
envelope of EXAFS spectrum (figure \ref{regM_FT}, middle). As seen, at small
$\alpha$ the noise dominates over the signal in the extended part of the
spectrum. The regularization parameter at which they match is the optimal
one, $\alpha_2$.

(3) For the Bayesian deconvolution it is necessary to find eigenvalues and
eigenvectors of a special symmetric matrix $g$. It turns out that the
determinant of this matrix varies with $\alpha$ over hundreds orders of
magnitude. At small $\alpha$'s the matrix is poorly defined, large $\alpha$'s
yield very large $\det(g)$ (figure \ref{regM}). Both cases give large
numerical errors because of ratios of very small or very large values in
calculations. As an optimal parameter we choose $\alpha_3=1.41$ at which
$\det(g)\sim1$.

The cases (1) and (2) require to determine the noise level, which demands
additional variables (for instance, the limiting frequencies of FT). The case
(3) does not explicitly concern the noise. Due to the dependence of
$\lg[\det(g)]$ on $\alpha$ appears to be linear, which readily allows one to
find the optimal parameter, the case (3) is more preferable from the
practical point of view. Below, for deconvolution of the real broadening
processes we use the optimal parameter $\alpha_3$.

\begin{figure}[!t]\begin{center}\includegraphics*{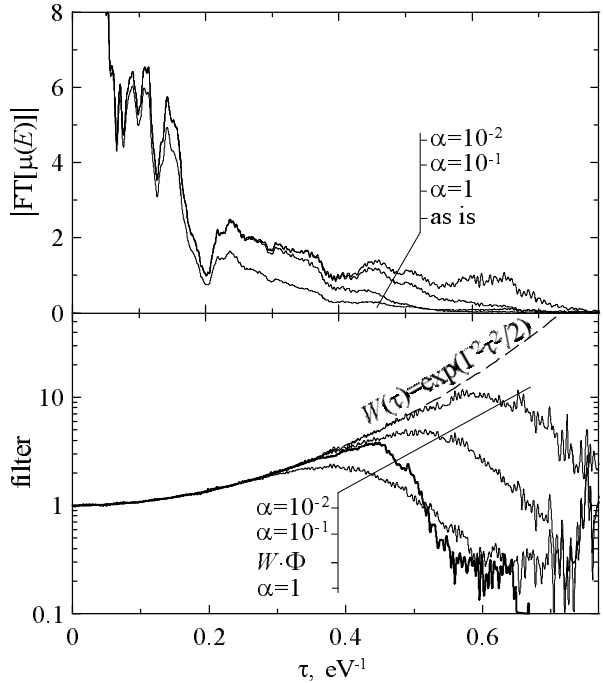}\end{center}
\caption{Module of the FT of initial $\mu_{\rm m}$ and deconvolved $\mu$ at
different $\alpha$. Below --- filters transforming $\mu_{\rm m}(\tau)$ to
$\mu(\tau)$ obtained after Bayesian deconvolution (thin solid lines), after
deconvolution based on the FT (dashed line), and after deconvolution based on
the combination of the FT and Wiener filtering (thick solid line).}
\label{rel2off}\end{figure}

It is of certain interest to compare the Bayesian method of deconvolution
with the widely known method combining optimal Wiener filtering and the
convolution theorem, where the conjugate variables of the FT are not $k$ and
$2r$ adopted in EXAFS but $E$ and $\tau$. According to the theorem, $\mu_{\rm
m}(\tau)=W(\tau)\cdot\mu(\tau)$, where for our model Gaussian response
$W(\tau)=\exp(\Gamma^2\tau^2/2)$. A simple back FT of the ratio $\mu_{\rm
m}(\tau)/W(\tau)$ will give the thought solution $\mu(E)$ but very noisy.
Therefore $\mu_{\rm m}(\tau)$ at large $\tau$ has to be smoothed. Figure
\ref{rel2off} shows module of the FT of the measured spectrum and of the
Bayesian deconvolved spectra. The latter are merged at
$\tau\lesssim0.25$\,eV$^{-1}$. In the bottom part of the figure the ratios
$|{\rm FT}\mu_{\rm deconv}|/|{\rm FT} \mu_{\rm m}|$ are shown for different
$\alpha$. As seen, the Bayesian deconvolution performs the effective
filtration of spectra with the limiting frequency $\tau_{\rm max}$ depending
on $\alpha$.

The optimal, in the least-square sense, Wiener filter is expressed as
\cite{NumRecipes}: $\Phi(\tau)=(1+|n(\tau)|^2/|{\rm FT}\mu_{\rm m}|^2)^{-1}$,
where $|n(\tau)|^2$ is the power spectrum of the noise replaced here by 0.01,
the mean value of $|\mu(\tau)|^2$ over the range $\tau>0.4$\,eV$^{-1}$. As
seen in figure \ref{rel2off}, the effective Wiener filter $W(\tau)\Phi$
transforming $\mu_{\rm m}(\tau)$ to $\mu(\tau)$ is close to the effective
filter of the Bayesian deconvolution with $\alpha=1\approx\alpha_3$. Notice,
however, that the limiting frequency for the estimation of noise power
spectrum was chosen rather arbitrarily.

In closing this section, it should be noticed that apart from the possibility
of determination of the deconvolution errors and the possibility of the
optimal regularization parameter choice, the Bayesian deconvolution has the
advantage of the capability to take into account {\em a priori} information
about the smoothness and shape of the solution (see details in
\cite{Klementev01}). In addition, in the Bayesian method the response
function $W(E-E')$ could be of more general form $W(E-E',E)$, which will be
useful for deconvolution of the instrumental broadening because monochromator
energy resolution depends noticeably on the angular position and, hence, on
the energy of the output x-ray beam.

\section{Applications of deconvolution}\label{instr}

We have seen that the Bayesian method proves to be effective for
deconvolution of EXAFS spectra, and the choice of the regularization
parameter appears to be irrelevant. Now we perform the deconvolution of
various types of broadening, for which purpose specify the corresponding
response functions $W(E-E',E)$.

\subsection{Instrumental broadening}

 The monochromator resolution is determined by the rocking curve
width $\delta\theta_B$ and by the vertical beam divergence
$\delta\theta_\perp$. For the monochromator Si(111) at $E=9$\,keV the rocking
curve width is $\delta\theta_B=32.4$\,$\mu$rad (FWHM) \cite{XOP}, the beam
divergence (LURE, D-21) $\delta\theta_\perp=150$\,$\mu$rad. Strictly
speaking, the resulting spectral distribution is given by the convolution of
rocking curve and the angular beam profile. But since
$\delta\theta_B\ll\delta\theta_\perp$, the energy selectivity is determined
by $\delta\theta_\perp$, namely: $\delta E_\perp/E
=\delta\theta_\perp\cot\theta_B= \delta\theta_\perp\sqrt{(2Ed/ch)^2-1}$,
where $\theta_B$ is Bragg angle, $d$ is Bragg plane spacing. Modelling the
spectral distribution by a Gaussian function, one obtains:
\[W(E-E',E)\propto\exp\left[-\frac{(E-E')^2}{2\sigma^2_\perp(E)}\right],
\sigma_\perp(E)=\frac{\delta E_\perp(E)}{2\sqrt{2\ln2}},\] where the
normalization constant must be calculated at each $E$ value. For our sample
spectrum, $\sigma_\perp(8850\,{\rm eV})=2.46$\,eV and
$\sigma_\perp(10500\,{\rm eV})=3.49$\,eV.

\subsection{Lifetime broadening}

For deconvolution of the lifetime broadening described by a Lorentzian
function $W(\Delta E)\propto[(\Delta E/\Gamma_K)^2+1]^{-1}$, we take as the
initial spectrum $\mu_{\rm m}$ the spectrum $\mu_{\rm instr}$ obtained after
the instrumental deconvolution. According to \cite{Krause79}, the width of Cu
$K$ level (FWHM) equals 1.55\,eV, from where $\Gamma_K=0.775$\,eV.

\subsection{Multielectron broadening}

There are certain difficulties in measuring XPS spectra near (and deeper) the
deepest atomic levels: the monochromatic x-ray sources of high energy are
required; for long enough spectra ($\sim100$\,eV) a photoelectron analyzer
with a broad energy window and long integration time are necessary.
Unfortunately, for lack of experimental XPS spectra in cuprates near Cu 1$s$
level, we can use a model representation of the response $W(\Delta E)$. For
that we take the estimations of position, intensity and width of the
secondary $KM_{23}$ excitation from \cite{DiCicco96:2}:
$E_{KM_{23}}-E_K=85$\,eV; $I_{KM_{23}}/I_K=0.03$; $\Gamma_{KM_{23}}=3$\,eV.
finally, for the broadening function we have: \[W(\Delta
E)\propto\frac{I_K\Gamma_K}{\Delta
E^2+\Gamma_K^2}+\frac{I_{KM_{23}}\Gamma_{KM_{23}}}{(\Delta
E-E_{KM_{23}})^2+\Gamma_{KM_{23}}^2}.\] Again, as the initial spectrum we
take the spectrum $\mu_{\rm instr}$.

\section{Discussion}\label{discussion}

\begin{figure}[!t]\begin{center}\includegraphics*{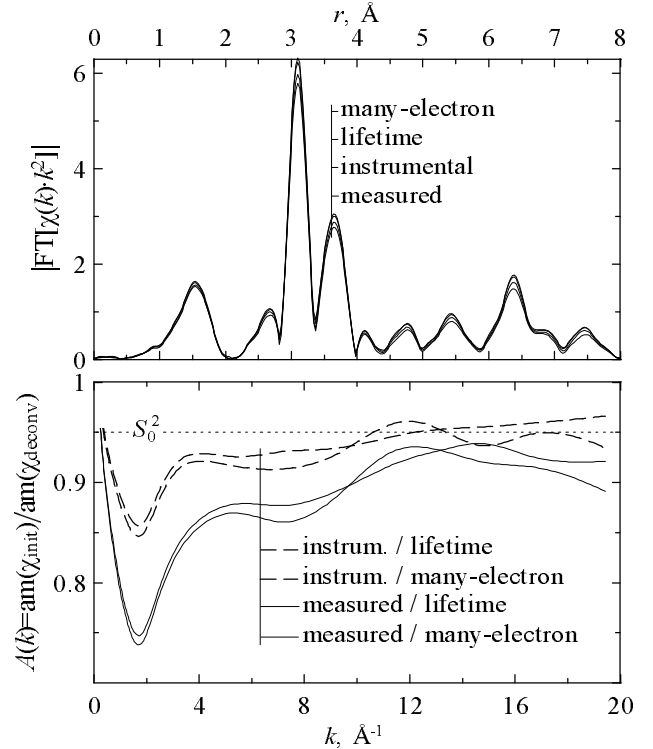}\end{center}
\caption{Module of the FT of various EXAFS functions: initial; obtained from
the instrumentally deconvolved absorptance; the latter was deconvolved with
the Lorentzian response and with the total multielectron response, these two
give the EXAFS FT's practically merged. Bottom: the ratio of amplitudes of
initial $\chi$ and deconvolved $\chi$. The $S_0^2$ value was calculated from
atomic overlap integrals.} \label{regFTS02}\end{figure}

With the specified response functions $W$, perform the Bayesian deconvolution
of the absorption coefficient at the optimal regularization parameter, then
construct the EXAFS function for which calculate FT and the amplitudes and
phases (see figure \ref{regFTS02}). Just as for the model response in Sec.
\ref{CompBackgr}, the deconvolution leads to the increase of EXAFS
oscillations. As appeared, the deconvolution has practically no influence on
the first FT peak originating from the shortest scattering path. It is clear
why: the oscillations corresponding to this peak are essentially wider then
the response $W$ (for these, $W$ is almost a $\delta$-function), and this is
more true for the extended part of a spectrum, due to the period of the
oscillations there is even longer (in $E$-space). Thus, it is in the extended
part where $\mu$, the solution of equation~(\ref{conv}), less differs from
$\mu_{\rm m}$.

In modern EXAFS spectroscopy the difference between amplitudes of
experimental and calculated spectra are taken into account by the reduction
factor $S_0^2$ which is either treated as a fitting parameter or estimated
from the relaxation of the core-hole as the many-electron overlap integral.
In many works this factor is considered to be independent from energy,
however, as noted in review by Rehr and Albers \cite{Rehr00}, it must be
path-dependent and energy-dependent. At the bottom of figure \ref{regFTS02}
we draw the ratios $A(k)$ of the amplitude of initial EXAFS spectrum to that
of the deconvolved one. Here, they were calculated relatively both $\chi_{\rm
m}$ and $\chi_{\rm instr}$. For comparison we show the factor $S_0^2$ as
computed by {\sc feff} code \cite{FEFF8}. At large $k$, where the noise
become comparable with the EXAFS signal, the ratios $A(k)$ have significant
errors. However, the general trend of the curves corresponds to the expected
one \cite{Rehr78}: $A(k)$ is minimal at intermediate EXAFS energies, while at
both low and high energy $A(k)$ reduces to unity. In addition, there are some
phase shifts between the initial spectrum and the deconvolved ones. But these
shifts are found to be quite small: less than 0.2\,rad at $k<4$\,\r{A} and
less than 0.1\,rad at $k>4$\,\r{A}.

The Lorentzian broadening of the EXAFS spectrum with a half-width $\Gamma$ is
similar to the effect of the imaginary part of the self-energy with ${\rm
Im\,}\Sigma=\Gamma$. The resulting reduction factors, i.e. the ratios the of
amplitudes calculated with and without the imaginary part, are analogous to
the reduction factors obtained by us relatively $\chi_{\rm instr}$:
$A(k)={\rm am}(\chi_{\rm instr})/{\rm am}(\chi_{\rm deconv})$. However, up to
now the reduction factors relatively {\em measured} EXAFS spectrum were
considered, $A(k)={\rm am}(\chi_{\rm m})/{\rm am}(\chi_{\rm deconv})$. As
seen (figure \ref{regFTS02}), these are the noticeably different quantities.
That is why for the correct analysis and comparison of spectra taken at
different experimental conditions, the instrumental deconvolution must be the
first step.

For our example spectrum and the chosen response functions, deconvolution of
the lifetime broadening and deconvolution of the multielectron broadening are
practically undistinguishable (figure \ref{regFTS02}, top), i.e. the
secondary weak peak in the excitation spectrum $W(\Delta E)$ has very little
effect. The main effect of using real excitation spectra is expected from the
presence and taking into account the plasmon losses which have a considerable
integral weight (figure \ref{Cicco}). Because of their very broad spectral
distribution, their effect consists in the change of the EXAFS spectrum as a
whole. In the present paper this contribution was not taken into account for
lack of appropriate experimental information.

Near the absorption edge, where the photoelectron kinetic energy is low, the
core-hole relaxation processes are of certain importance for the
photoelectron propagation. Here we do not consider the validity of the
neglect of this effect, but refer to the review \cite{Rehr00}.

\section{Conclusion}\label{conclusion}

To take into account the many-electron effects, there exist, in principle,
two approaches: (a) to include into a one-electron theory relevant amendments
or (b) to extract a one-electron absorptance from the total one and to use
then a pure one-electron theory. The first, traditional, approach invokes
semi-empirical rules, but not {\em ab initio} calculations, to construct the
exchange correlation part of the scattering potential, with the empiricism
being based on the comparison with experimental spectra {\em already
broadened}. In the present paper we have shown the principle way for the
second approach, using the solution of integral convolution equation, the
kernel in which is the excitation spectrum measured in XPS spectroscopy.
Notice, that owing to the specific way of the structural information
extraction from the EXAFS spectra, in which an isolated signal in $r$-space
or a filtered signal in $k$-space is used, we have not committed a sin
against the fact that the integral convolution equation is an ill-posed
problem, because from the infinitely large number of solutions it is possible
to determine an optimal range, also infinitely large, of solutions any one
from which is appropriate.

Because of some technical difficulties, it is impossible so far to measure
XPS spectra near deep core levels. Therefore, the desirable pure one-electron
absorptance is unavailable. Nevertheless, as we have shown, it is possible to
perform an accurate instrumental deconvolution and deconvolution of the
lifetime broadening. These procedures make the comparison between calculated
and experimental spectra more immediate and the final results of EXAFS
spectroscopy more reliable.

All the stages of EXAFS spectra processing including those described here are
realized in the freeware program {\sc viper} \cite{VIPER}.

\acknowledgements The example spectrum was measured by Prof. A.~P.
Menushenkov. The author wishes to thank Dr. A.~V. Kuznetsov for many valuable
comments and advices. The work was supported in part by RFBR grant No.
99-02-17343.


\end{document}